\documentclass[10pt,conference]{IEEEtran}
\usepackage{algorithmic,algorithm}
\usepackage[a4paper,bindingoffset=0.2in,left=0.667in,right=0.68in,top= 0.75in,bottom=1.1in,footskip=.25in]{geometry}
\usepackage{amsmath}
\usepackage{amssymb,amsmath}
\usepackage{graphicx}
\usepackage{epsfig}
\usepackage{grffile}
\usepackage{balance}
\usepackage{balance,color}
\usepackage{algorithmic,algorithm}
\providecommand{\keywords}[1]{{\textit{Index Terms}}}
\abovecaptionskip 
\belowcaptionskip 

\begin{document}

\title{Edge Computing For Smart Health: Context-aware Approaches, Opportunities, and Challenges}
\author{Alaa Awad Abdellatif$^{*\dag}$, Amr Mohamed$^{*}$, Carla Fabiana Chiasserini$^{\dag}$, Mounira Tlili$^{+}$, and Aiman Erbad$^{*}$\\
$^*$Department of Computer Science and Engineering, Qatar University \\
$^\dag$Department of Electronics and Telecommunications, Politecnico di Torino  \\
$^+$Department of Computer Science, Carnegie Mellon University \\
E-mail: \{aawad, amrm, aerbad\}@qu.edu.qa, carla.chiasserini@polito.it, Mounirastlili@gmail.com \\
	%
}
\maketitle

\begin{abstract}
Improving efficiency of healthcare systems is a top national interest worldwide. However, the need of delivering scalable healthcare services to the patients while reducing costs is a challenging issue. Among the most promising approaches for enabling smart healthcare (s-health) are edge-computing capabilities and next-generation wireless networking technologies that can provide real-time and cost-effective patient remote monitoring. 
 In this paper, we present our vision of exploiting multi-access edge computing (MEC) for s-health applications. We envision a MEC-based architecture and  discuss the benefits that it can bring to realize in-network and context-aware processing so that the s-health requirements are met. 
 We then present two main functionalities that can be implemented leveraging such an architecture to provide efficient data delivery, namely, multimodal data compression and edge-based feature extraction for event detection. The former allows efficient and low distortion compression, while the latter  ensures high-reliability and fast response in case of emergency applications.   
 Finally, we discuss the main challenges and opportunities that edge computing could provide and possible directions for future  research.
\end{abstract}
\begin{IEEEkeywords}
Edge computing, smart health, Internet of Medical Things (IoMT), context-aware processing, deep learning.
\end{IEEEkeywords}

\section{Introduction\label{sec:Introduction}}

The evolution of computational intelligence and Internet of Medical Things (IoMT), along with the advances of next-generation wireless technologies, has boosted the development of traditional healthcare processes into smart-healthcare services. 
 Smart-health (s-health) can be considered as the context-aware evolution of mobile-health, leveraging wireless  communication technologies to provide healthcare stakeholders with innovative tools and solutions that can revolutionize service provisioning. 
In particular, s-health enables remote monitoring, where patients and caregivers  leverage mobile technologies to provide information about their health remotely -- a service that is expected to  reduce hospitalization considerably and allow for timely delivery of healthcare services to remote communities at low costs. 

S-health systems will also leverage various wireless sensors, cameras, and controllers, which permit patients' automatic identification and tracking, correct drug–patient associations, and intensive real-time vital signs monitoring  for early detection of clinical deterioration (e.g., seizure detection, heart failure, etc.). All these {\em things} will  report an impressive amount of data that need to be transported, swiftly processed, and stored, while ensuring privacy protection.  
Given these requirements, the conventional cloud computing paradigm becomes unsuitable for s-health, since a centralized approach cannot provide  a sufficiently high level of scalability and responsiveness, and it will impose while an exceedingly heavy traffic load to communication networks.   
A new approach has therefore emerged, known as Multi-access Edge Computing (MEC), 
defined as the ability to process and store data at the edge of the network, i.e., in the proximity of the data sources. 
The advantage of MEC in a smart heath environment is multifold as it can provide short response time, decreased energy consumption for battery operated devices, network bandwidth saving, as well as secure transmission and data privacy \cite{alaa_IoTjournal}. Furthermore, it can be applied to various network scenarios, including cellular, WiFi and fixed access technologies. 
 This paper paves the way for MEC usage in smart heath environment through answering the following questions:
\begin{itemize}
	\item What are the motivations and main expected benefits of leveraging the MEC architecture in s-health systems?
	\item What are the s-health requirements, solutions of MEC, and open challenges?
\end{itemize}

 In what follows, Section \ref{sec:why}  introduces a MEC-based system architecture that meets the 
s-health requirements, highlighting the benefits of pushing data processing and storage toward the data sources.   Section \ref{sec:Cases} presents context-aware solutions  for implementing multimodal data compression, in-network processing, and event-detection at the edge.  
Section \ref{sec:Opportunities} then discusses some challenges that MEC poses and further opportunities that such a paradigm offers. 
Finally, Section \ref{sec:conclusion} concludes the paper. 

\begin{figure*}[htp]
	\centering
		\scalebox{3.78}{\includegraphics[width=0.27 \textwidth]{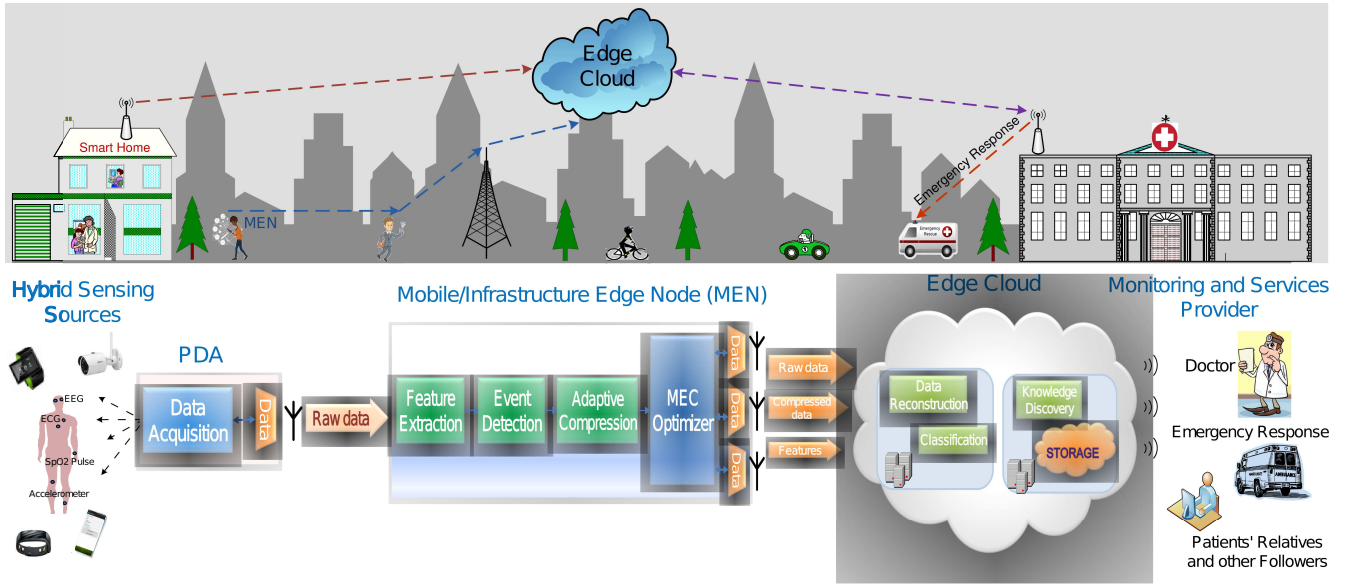}}
	\caption{Proposed smart health system architecture. }
	\label{fig:system_model}
\end{figure*}
\section{MEC-based Architecture for Smart Health \label{sec:why}}
We now give a brief description of the proposed MEC-based architecture for e-health applications, and discuss the benefits that it offers to s-health systems. 

\subsection{MEC-based S-Health  Architecture \label{sec:Benefits}} 

The proposed system architecture, shown in Figure~\ref{fig:system_model}, stretches from the data sources located on or around  patients to the service providers. It contains the following major components: 

\textbf{Hybrid sensing sources:} A combination of sensing devices attached/near to the patients represent the set of data sources. Examples include: body area sensor networks (including implantable or wearable medical and non-medical sensors), IP cameras, smartphones, and external medical devices. All such devices are leveraged for monitoring patients' state within the smart assisted environment, which facilitates continuous-remote monitoring and automatic detection of emergency conditions. These hybrid sources of information are attached to a mobile/infrastructure edge node to be locally processed and analyzed before sending it to the cloud (see Figure~\ref{fig:system_model}).   

 \textbf{Patient Data Aggregator (PDA):} Typically, the wireless Body Area Network (BAN) consists of several sensor nodes that measure different vital signs, and a PDA which aggregates the data collected by a BAN and transmits it to the network infrastructure. Thus, the PDA is working as a communication hub that is deployed near to the patient to transfer the gathered medical data to the infrastructure. 

\textbf{Mobile/Infrastructure Edge Node (MEN):}  Herein, a MEN implements intermediate processing and storage functions between the data sources and the cloud.  
The MEN fuses the medical and non-medical data from different sources, performs in-network processing on the gathered data, classification and emergency notification, extracts information of interest, and 
forwards the processed data or extracted information to the cloud. 
Importantly, various healthcare-related applications (apps) can be implemented in the MEN, e.g., 
for long-term chronic disease management. Such apps can help patients to actively participate in their treatment and to ubiquitously interact with their doctors anytime and anywhere.  
 Furthermore, with a MEN running specialized context-aware processing, various data sources can be connected and managed easily near the patient, while optimizing data delivery based on the context (i.e., data type, supported application, and patient's state) and wireless network conditions.  


\textbf{Edge Cloud:} It is a local edge cloud where data storage, sophisticated data analysis methods for pattern detection, trend discovery, and population health management can be enabled. 
An example of the edge cloud can be a hospital, which monitors and records patients' state while providing required help if needed.  

\textbf{Monitoring and services provider:}  A health service provider can be a doctor, an intelligence ambulance, or even a patient's relative, who provides preventive, curative, emergency, or rehabilitative healthcare services to the patients.         

\subsection{Benefits for s-health \label{sec:Benefits}}

Given the characteristics and requirements of e-health applications, Table \ref{tab:RelWork} summarizes some of the e-health systems that can benefit from the above architecture. It is not the objective of this paper to provide an in depth technical comparison on the different proposed e-health systems. However, we investigate the practical benefits of leveraging MEC in such systems. In what follows, we will discuss the advantages of the proposed MEC architecture in the light of these systems.  

\begin{table*}[htp]
	\centering
		\caption{Summary of the e-health relevant systems.}
	  \label{tab:RelWork}
		\begin{tabular}{|c|c|c|c|c|} 
			\hline 
		\textbf{Application} &\textbf{Collected Data} &\textbf{Description} & \textbf{Limitations} & \textbf{MEC Benefits}\\
		\hline
	 	Cardiac disorder &Electrocardiography& Heart monitoring system is developed &All data processing &Data reduction  \\
		detection \cite{cardiac}&(ECG) &for detecting status of the patient &tasks are performed &and BW saving  \\
		& &and sending an alert message&at a web server & \\
		& & in case of abnormalities& & \\
		& &\textit{Requirements}: long lifetime for & & \\
		& &the battery-operated devices& & \\		
		\hline
		Remote Cardiac &Heart rate&A location based real-time Cardiac &Fewer number of &Location  \\	
		monitoring \cite{CardiacPatients}&blood pressure&monitoring system is developed&subjects participated &Awareness and \\
		&body temperature&\textit{Requirements}: long lifetime for &in the experiments&energy saving \\
		& &the battery-operated devices& & \\	
		\hline
		Detection of Ischemia &ECG and Electronic &Presenting different methods leveraged &Majority of the reviewed &Data reduction \\
		and Myocardial &Health Records (EHR)&ECG signal with EHR information &literature did not exploit &and BW saving\\
		Infarction \cite{Myocardial}& &to detect Ischemia and MI&contextual information & \\
		& &\textit{Requirements}: low computational & & \\
		& &complexity& & \\
		\hline
	Parkinson’s disease (PD)&Voice signal&A PD monitoring system over the cloud &All data processing &Data reduction \\
	detection \cite{Parkinson}& &is proposed using feature selection and &tasks are performed &and BW saving  \\
	& &classification of a voice signal&at the cloud&  \\
		& &\textit{Requirements}: Reliability and high & &\\
		& &classification accuracy& &\\
		\hline
		Contactless heart rate &Heart rate&Heart rate measurement from facial &Illumination variance, &Data reduction  \\
		measurement \cite{facialvideo}& &videos using digital camera sensor&motion variance, and &and BW saving\\
		& &\textit{Requirements}: Reliability and high &motion artifacts&\\
		& &measurement accuracy& &\\
	\hline
	Prediction of &ECG&Leveraging point process analysis of &Using single-channel &Low latency \\
	Bradycardia in preterm & &the heartbeat time series to predict &ECG data to predict	 & \\
	infants \cite{Bradycardia}& &infant Bradycardia prior to onset&Bradycardia& \\
	& &\textit{Requirements}: Fast prediction of & & \\
		& &emergency situations& & \\
	\hline
	Real-time epileptic &Electroencephalography&Automatic epileptic seizure detection &Requiring large amount &Low latency \\
	seizure detection \cite{Seizure2017}&(EEG)&system using wavelet decomposition &of data for training& \\
	 & &is proposed&to improve specificity & \\
		& &\textit{Requirements}: Fast seizure detection&of the detector& \\
		\hline	
	ECG change &ECG&A centralized approach for the detection &Using one type of	data &Low latency \\
	detection \cite{heart_attack} & &of abnormalities and intrusions in &for detecting abnormality & \\
	 & &the ECG data is developed&and emergency situations& \\
	& &\textit{Requirements}: Fast detection of & & \\
		& &abnormalities& & \\
	\hline
	Remote monitoring of &Pulmonary Function &Real-time tracking system of chronic &Relying on one type	&Low latency \\
	chronic obstructive &Test (PFT)&pulmonary patients comfortable in &of data& \\
	pulmonary \cite{Pulmonary}& &their home environment is developed&	& \\
		& &\textit{Requirements}: Fast detection of & & \\
		& &abnormalities& & \\
		\hline	
		\end{tabular}
\end{table*}

\subsubsection{Monitoring systems using wearable devices\label{sec:Wearable}} 
Heart monitoring applications are the most common type of remote monitoring applications. Monitoring vital signs related to the heart reveals many types of diseases, e.g., Cardiac arrhythmia, chronic heart failure, Ischemia and Myocardial Infarction \cite{cardiac}\cite{CardiacPatients}\cite{Myocardial}. In \cite{cardiac}, authors present a real-time heart monitoring system, where the extract medical data of the patients are transmitted to an Android based listening port via Bluetooth. Then, this listening port forwards these data to a web server for processing. Also, \cite{CardiacPatients} exploits Android smartphone to gather patient’s information from wearable sensors and forward it to a web portal in order to facilitate the remote cardiac monitoring. However, in these systems, the smartphone is used only as a communication hub to forward collected data to the cloud. 
 Hence, continuous data transmission is not viable due to the high energy toll it implies. 
 The advantages of implementing the proposed MEC architecture in such systems are twofold. First, energy saving can be significantly increased by carefully managing the devices operational state and their data transfer at the MEN. In addition, data compression as well as  the proximity between sensors and MEN further reduce the  energy consumption due to data transmission \cite{ALaa2017Elsevier}\cite{Alaa_SS}.    
Second,   the network edge can be fruitfully exploited to extract context information and apply localization techniques,  
which allows matching the patient’s geographical position with the nearest appropriate caregivers (e.g., hospital or ambulance).  

\subsubsection{Contactless monitoring systems\label{sec:Contactless}}
Along with the evaluation of remote sensing, contactless monitoring has attained much focus recently. The main motivation of using contactless sensors is enabling ordinary life as much comfortable as possible to all patients, since the patients are required only to be present within a few meters from the sensors \cite{Parkinson}. Heart rate measurement from facial videos using digital camera sensors is one of the rapidly growing directions to extract physiological signals without affecting patient's activities \cite{facialvideo}. However, transmitting large volumes of data generated from these camera sensors using conventional cloud-based architecture is not advisable and may deem some of these applications impractical given the limited bandwidth availability.  For instance, the amount of digital data generated from a single-standard camera can reach to $40$ GB per day. 
Accordingly, processing, compressing, and extracting most important information from the gathered data at the MEN greatly reduce the amount of data to be transferred toward the cloud, hence the bandwidth consumption, and even makes it possible to store the data locally. 

\subsubsection{Disorder prediction/detection systems\label{sec:Prediction}}
 One of the promising applications of s-health, is the predictive monitoring of high-risk patients. The aim of these techniques is improving prediction/detection of the emergency to implement preventative strategies for reducing morbidity and mortality associated with high-risk patients. For instance, \cite{Bradycardia} presented a simplistic framework for near-term prediction of Bradycardia in preterm infants using statistical features extracted from ECG signal. Also, \cite{Seizure2017} proposed a quick seizure detection algorithm using fast wavelet decomposition method.   
In such real-time prediction/detection systems, the swift delivery of data to the server is a necessity.  
In many cases, this requires that data are analyzed and even a diagnosis is made as close as possible to the patient. 
However, detecting the changes of the physiological signals (e.g., changing in ECG values) in continuous health monitoring systems is not an easy task. It can be an indication for an emergency situation (e.g., occurrence of a heart attack) \cite{heart_attack}\cite{Pulmonary}.  This abnormality detection task becomes even more challenging during wireless communication transfer of patient’s data to the cloud due to the erroneous communication and security attacks that could introduce errors or makes changes in the patient’s data. 
 Hence, quick detection of the changes in the gathered medical data at the MEN is essential for real-time abnormal event detection. 
In a nutshell, the implementation of MEC architecture addresses all these issues, and the ability of the MEN to perform event detection/prediction fulfills these requirements even in the case of emergency applications. 

\section{Implementing the Edge Node Functions \label{sec:Cases}}

The ultimate goal of our MEC architecture is to fulfill the different requirements of e-health applications mentioned above and enable s-health services through implementing the following main functionality at the network edge:
\begin{itemize}
\item data compression, in order to reduce energy and bandwidth consumption in the case of health monitoring systems;
\item feature extraction and classification, in order to ensure high-reliability and fast response time in disorder prediction and detection.
\end{itemize} 
Below, we present how the above functionality can be implemented at the MEN and highlight the benefits that the MEC architecture can bring. 

\subsection{Multimodal data compression using deep learning\label{sec:Multimodal}}

The conventional approach used for health monitoring, i.e., transmitting the entire medical data wirelessly to the cloud, implies the transfer of a massive amount of data. For instance, in brain disorder monitoring systems, EEG, Electromyography (EMG), and Electrooculography (EOG) data need to be stored and accessed remotely, along with video recording patient's activities, in order to correlate the patient's activities with her EEG pattern. This would result in generating 8-10\,GB per patient per day.  
A promising methodology to deal with this issue in s-health systems is to perform local in-network and data-specific compression on the gathered data before transmission, while taking into account the applications' requirements and the characteristics of the data. 

Here, we consider the EEG-EOG monitoring system as a case study and present an efficient technique that deals with multimodal data, as required by s-health applications. In particular, we use the multimodal dataset in \cite{DeepLearning2018}, which contains EEG and EOG signals of $32$ people, who volunteered for this experiment, watching to 40 music videos.  

Our solution leverages deep learning, which is  a good candidate for multimodal data compression due to its ability to efficiently exploit, not only the intra-modality correlation, but also inter-correlation among different modalities. Specifically, we use Stacked Auto-Encoders (SAE), i.e., a 
special type of neural networks allowing for the hierarchical extraction of data representation \cite{DeepLearning2018}. 
SAE consists of two main layers: (i) the encoding layers where the data features are extracted, 
 and (ii) the decoding layers where the signal is reconstructed from the data coming from the encoding layers.
In our case, we implement the encoding layers at the MEN, while the decoding layers are placed at a server in the cloud.  Our key idea is to progressively  reduce the number of neurons in each of the encoding layers at the MEN, and make the neural network learn from compressed version of the data. As a result, through the last encoding layer at the MEN (i.e., the one with the least number of neurons), we obtain a set of features that are a compressed representation of the initial data. 
In summary, at the MEN, our  SAE encoder converts the input data $x$ into the compressed data $z$, provided by the last encoding layer. At the server side (in the cloud), the SAE decoding layers will obtain the reconstructed data $\tilde{x}$, using the compressed representation $z$. The compressed and reconstructed signals can be written as:
\begin{equation} 
 z = Wx+b 
\label{z}
\end{equation}
\begin{equation} 
 \tilde{x} = \tilde{W}z+\tilde{b}. 
\label{z}
\end{equation}
where $W$ and $\tilde{W}$ are the encoder and decoder weights matrices, respectively, while $b$ and $\tilde{b}$ are the bias vectors. 

The objective of SAE is to find the optimal configuration of the weight matrices and bias vectors that minimize the reconstruction error. In our case, instead, we first set the number of neurons in the last layer at the encoder, according to the desired compression ratio. Then we optimize the  number of neurons to be placed in each of the other encoding layers.  Finally, by training the neural network, we determine the optimal  weight matrices and bias vectors. 
We remark that, although the network training is a computational expensive task, it can be conducted offline at the server side. Then such a configuration can be sent and used at the MEN for on-line data compression, thus  leading to low-complexity, on-line data compression and transfer.     

\begin{figure}[htp]
	\centering
		\scalebox{1.66}{\includegraphics[width=0.27 \textwidth]{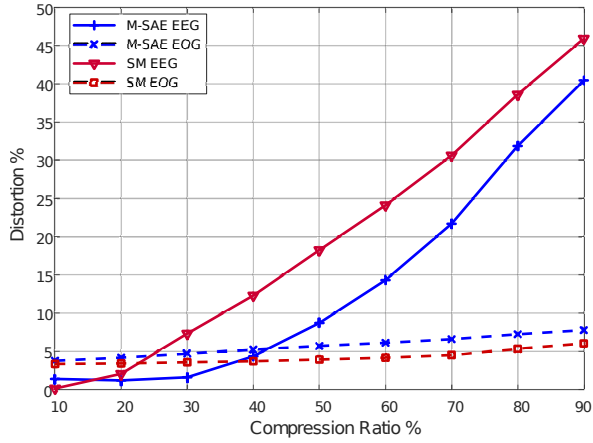}}
	\caption{Signal distortion as a function of the compression ratio for EEG and EOG signals, using M-SAE and SM compression schemes.}
	\label{fig:multimodal}
\end{figure}

The advantages of multimodal data over single modality compression are twofold. First, we can account for inter-modality correlation during compression, which results in a lower distortion while reconstructing the signal. Second, by  concatenating the different modalities (i.e.,  EEG and EOG signals), it enables encoding the modalities in a single-joint representation (i.e., the single stream $z$).  
Figure~\ref{fig:multimodal} compares the proposed multimodal SAE (M-SAE) with the Single Modality (SM) compression scheme, which compresses each signal separately using SAE. As the compression ratio varies, we observe that multimodal SAE allows for up to $50\%$ reduction in EEG distortion with respect to SM, while EOG distortion increases by just 2\%. 
\subsection{Edge-based feature extraction and classification \label{sec:Classification}}    

Many neurodegenerative diseases detection methods, such as Parkinson’s, Epilepsy, Alzheimer’s, and Huntington’s, have been reported in the literature based on extracting some features from the patients' vital signs, voice, or captured videos. Such features are used to differentiate a potential patient from a healthy person, or to identify emergency situations. For instance, \cite{Parkinson} proposes a method to detect Parkinson's disease (PD) leveraging certain features of the voice signal using cloud computing. Specifically, at the cloud server, the voice signal features are extracted and used for classification; the results are then sent to registered doctors for  proper action.  

In our study, we focus on epileptic seizure detection and show the advantages of implementing feature extraction and classification at the MEN for efficient transmission and fast detection of abnormalities. We assume that the MEN gathers EEG data from the patient using an EEG Headset, processes the data, and forwards them to the cloud. We now exploit the EEG dataset used in \cite{alaawts2018}, which comprises three classes of data, in the following denoted by A, B, and E, respectively.  Each class contains 100 EEG records corresponding to different patients. Each record includes samples collected for  23.6 seconds at a 173.61\,Hz rate. Sets A and B represent seizure-free subjects with eyes opened (A) and closed (B), respectively, while set E contains data related to epileptic patients. 

\begin{figure}[htp]
	\centering
		\scalebox{1.8}{\includegraphics[width=0.27 \textwidth]{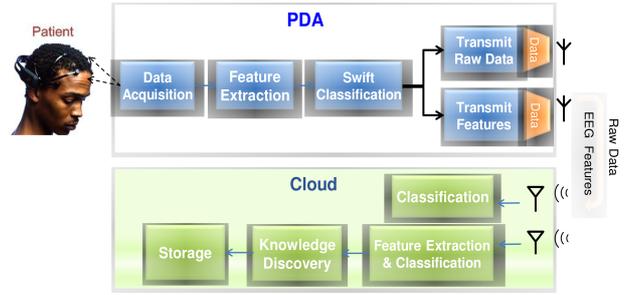}}
	\caption{Efficient class-based data transmission for s-health systems.}
	\label{fig:Event_detection}
\end{figure}

Using such data, we first perform feature extraction and classification at the MEN. Then, depending on whether a seizure event was detected through classification, the system sends to the cloud server the all data, or only the computed features. Figure \ref{fig:Event_detection} summarizes the proposed key concept. 
Below, we describe an efficient technique for feature extraction in Sec. \ref{sec:FeatureExtraction}, then we address event detection and classification in Sec. \ref{sec:applicationspecific}.  

\subsubsection{Feature extraction \label{sec:FeatureExtraction}}

In order to carry out the first step of our procedure, two approaches can be implemented: time-domain and  frequency-domain feature extraction. Herein, we consider the frequency-domain approach due to its insensitivity to signal variations resultant from  electrode placement. By transforming the gathered EEG data into the frequency domain, the normal/abnormal EEG classes under study exhibit different mean, median, and amplitude variations. Also, Root Mean Square (RMS) and Signal Energy (SE) are good signal strength estimators in different frequency bands. Hence, to distinguish between seizure and non-seizure events we select the following five Frequency Features (FF): mean ($\mu$), median (M), peak amplitude (P), RMS, and SE. 

\subsubsection{Event-detection at the edge \label{sec:applicationspecific}}

The second step in our procedure consists in developing a reliable, edge-based classification technique for seizure
detection leveraging the extracted features \cite{alaa_ExpertSystems2018}.  
A number of machine learning techniques, including supervised, unsupervised and reinforcement learning, have been investigated for the purpose of classification, for a variety of applications. In a nutshell, supervised learning algorithms leverage a labeled training data set to learn the relation between inputs and outputs. In contrast,  unsupervised learning algorithms classify the provided data sets into different clusters by discovering the correlation between input samples. The third category includes reinforcement learning algorithms and exploits online learning, which involves the exploration of the environment and the exploitation of current knowledge, in order to classify the data \cite{Machine_Learning_Wireless_Sensor2014}.  However, some important issues arise when machine learning techniques are applied to s-health: (i) an optimal trade-off between  algorithms computational complexity and classification accuracy should be established, (ii) sufficiently large datasets have to be considered, in order to ensure high accuracy, (iii) a mathematical formulation of the learned model, as well as full control over the knowledge discovery process, is hard to obtain.

In the considered case study, we define an IF-THEN classification rule using generated FF to detect abnormal variations in sensed EEG data due to seizure. Thus, the status of the patient, $S$, is given by:
\begin{equation}
S \mathord{=} \begin{cases}
     \text{Normal}&  \text{if}\, \mu\mathord{+}M\mathord{+}P\mathord{+}RMS\mathord{+}SE \leq \gamma \\
     \text{Seizure}& \text{if}\, \mu\mathord{+}M\mathord{+}P\mathord{+}RMS\mathord{+}SE > \gamma
\end{cases}\nonumber
\label{class_rule}
\end{equation}
 where $\gamma$ is the classification threshold obtained during the offline training phase.  
Thus, leveraging the proposed low-complexity classifier, a quick emergency notification system can be implemented at the edge to notify patient's caregivers in case of emergency, as well as doctors at the remote site.  

\begin{figure}[htp]
	\centering
		\scalebox{1.7}{\includegraphics[width=0.27 \textwidth]{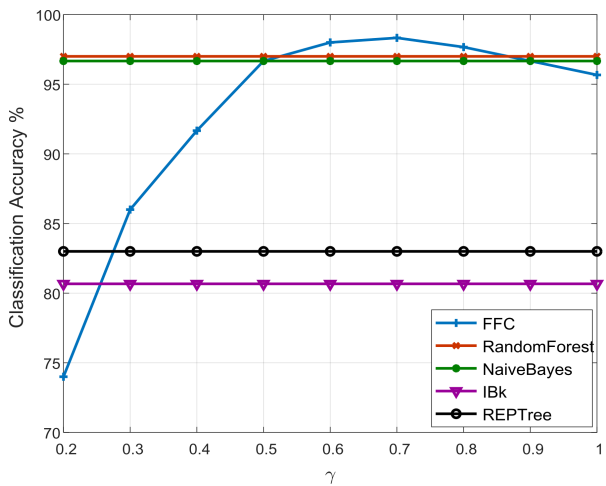}}
	\caption{Comparison of the proposed FFC technique with respect to RandomForest, NaiveBayes, IBk, and REPTree algorithms, in terms of classification accuracy, with varying $\gamma$.}
	\label{fig:Classification}
\end{figure}

 In Figure~\ref{fig:Classification}, we compare the accuracy of the proposed Frequency Features Classifier (FFC) against that of different machine learning approaches, including random decision forests (RandomForest), Naive Bayes (NaiveBayes), k-Nearest Neighbors (IBk), and classification/regression trees (REPTree). Each of these classifiers is run using the default algorithm configuration in WEKA explorer with 5-fold cross-validation \cite{weka}. 
In FFC, when $\gamma$ is small, most of the obtained statuses will be Seizure, while at high values of $\gamma$, most of the obtained statuses will be Normal. In both cases, our classifier cannot accurately differentiate between the patients’ classes. In the
middle region, when $\gamma$ ranges between 0.5 and 0.8, our classifier can discriminate between different classes yielding a high accuracy. Notably, for $\gamma$ ranging between 0.5 and 0.8, our classifier outperforms other solutions, achieving 98.3\% accuracy for seizure detection for $\gamma=0.7$.

\begin{figure}[htp]
	\centering
		\scalebox{1.7}{\includegraphics[width=0.27 \textwidth]{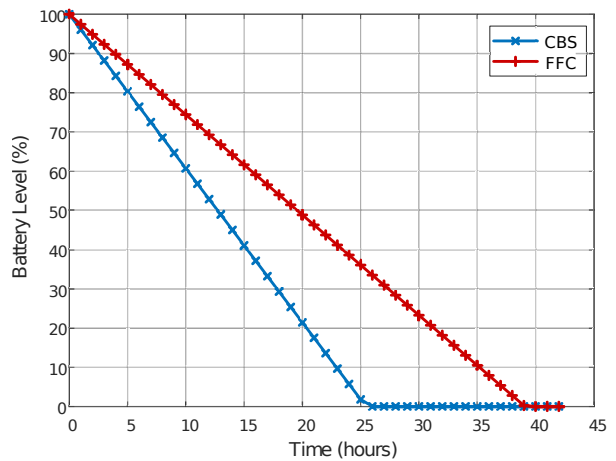}}
	\caption{Comparison of the proposed FFC technique with respect to CBS scheme, in terms of battery lifetime.}
	\label{fig:Battery_Level}
\end{figure}

 Figure~\ref{fig:Battery_Level} assesses the performance of the proposed class-based data transmission (CDT) scheme, in terms of MEN's battery lifetime, comparing with cloud-based system (CBS). CDT refers to the proposed scheme where the MEN locally classifies the acquired data to decide whether the all data, or only the computed features, should be sent to the cloud, while CBS  refers to a traditional system where the MEN is used only as a communication hub to forward all acquired data to the cloud. 
Herein, we used a Samsung Galaxy S4 smartphone as a MEN, which is connected to both a data emulator and server via WiFi. Interestingly, Figure~\ref{fig:Battery_Level} shows that CDT can improve the MEN's battery lifetime by 50\% with respect to CBS.  

\section{Challenges and Opportunities \label{sec:Opportunities}}

In this section, we discuss three main challenges and opportunities that characterize MEC-based s-health systems and represent interesting lines for future research. 

\subsection{privacy and security}

Great potential of s-health system can only be achieved if individuals are confident about the privacy of their health-related information and providers are confident about the security of gathered data. However, ensuring privacy and security is not straightforward. Wireless medical devices are typically susceptible to various types of threats, such as patient tracking and relaying, as well as denial of service attacks, which violate confidentiality and integrity of the devices. Data processing algorithms and data storage may also be subject to attacks. Below, we discuss some challenges and opportunities that MEC poses in this respect. 

First is the ownership of the collected data from the patients. Storing the data at the patients' proximity, where it is collected, and enabling the patients to fully own the data is a better solution for privacy protection. Also, the patient will be able to control if the data should be stored at the edge or transmitted to the cloud after removing or hiding some of the private information from the data.  

 Second is the trade-off between increasing security level and QoS. Increased security through strong cryptographic algorithms or effective key management schemes \cite{KeyManagement2009}, adds more processing and additional overhead at the edge, which may have a significantly adverse impact on QoS, especially for real-time applications with strict delay and throughput requirements.  
This imposes an essential need to design joint QoS and security mechanisms for s-health applications that maximize QoS while meeting the application security requirements.

\subsection{Collaborative edge}
 Healthcare requires data sharing and collaboration among different stakeholders in multiple domains. However,  sharing of data owned by a stakeholder  rarely happens due to privacy concerns and the high cost of data transfer. 
 { In this context, collaborative edge, which connects the edges of multiple stakeholders that are geographically distributed (such as hospitals, centers for disease control and prevention, pharmacies, and insurance companies),  is beneficial in threefold. }  First, it provides distributed data sharing among different stakeholders at low cost,  thanks to  computation and processing at the participant edges.
 Second, in the case of remote monitoring, it enables patients to forward their medical data to the cloud through other users/edge nodes. This also improves spectrum and energy efficiency and allows data transferring even in geographically remote areas by exploiting D2D data transfer  \cite{ALAA_D2D}.  
Third, it enables a patient's  edge node to directly connect to the nearest hospital's edge in the proximity for continuous monitoring, without the need of going through the cloud. This helps to increase monitoring efficiency, reduce energy consumption and operational cost, as well as improve high-quality services. 

\subsection{Combining heterogeneous sources of information}

 Various sources of information are used in S-health systems for efficient monitoring, hence, leveraging  advanced multimodal data processing techniques for combining these sources of information at the edge is a promising trend toward automating supervision and remote monitoring tasks.      
However, several challenges remain open when it comes to the s-health systems with hybrid sensing sources. First, in terms of multiple modalities, it is not straightforward to incorporate and transmit multiple data streams in s-health systems, where power consumption is a limiting factor; indeed,  transmission of highly informative biosignals (e.g., EEG, EMG, and electrocardiogram) is an energy hungry process for battery-operated devices. Second, signals artifacts arise from internal sources, e.g., muscle activities and movements, as well as from external sources related to noise, interference, and signals offset, which have critical implications on data quality \cite{Artifact_Removal2012}.   

 In this context, adopting a MEC-based s-health system architecture would be beneficial in two ways.  First, it permits to address system complexity associated with such heterogeneous and variable data-stream inputs. This is done through implementing multimodal in-network processing techniques that yield the correlation between different modalities, in addition to the temporal correlation within each modality. Moreover, a MEC-based architecture enables extracting high level application-based features at the edge rather than the cloud. By doing so, a MEN can send a limited number of the extracted features, or the obtained correlations, instead of transmitting either the original or the compressed data.   
Second, advanced signal processing for artifact removal can be incorporated at the edge, in order to improve signals quality before  transmission.

\section{Conclusion\label{sec:conclusion}}

In this paper, we presented our vision of an s-health system leveraging the multi-access edge computing paradigm. Such an approach can indeed boost the system performance by efficiently handling the enormous amount of data generated by sensors and  personal as well as medical devices at the edge of the network, and addressing the limited energy capabilities of such devices. 
In particular, edge-based processing like compression and event detection can greatly reduce the amount of data transferred toward the cloud, thus removing one of the major bottlenecks in s-health systems. Furthermore, processing data at the edge will ensure better user privacy than when raw data is uploaded to the cloud. 
Accordingly, we recommend that wireless network components, characteristics of the acquired data, and high level requirements of the considered application should be integrated in order to provide sustainable and high-quality services for s-health systems.   
  In this context, we identified some computing tasks that can be implemented at the edge and presented effective approaches to implement them, so as to ensure short response time, efficient processing and minimal energy and bandwidth consumption. Finally, we highlighted some challenges and opportunities of edge computing in the s-health field that are worth further research.
	
\balance
\section*{Acknowledgment}
This work was made possible by GSRA grant \#  GSRA2-1-0609-14026  from the Qatar National Research Fund (a member of Qatar Foundation). The work of Amr Mohamed and Alaa Awad Abdellatif is partially supported by NPRP grant \# 8-408-2-172. The findings  achieved  herein are solely the responsibility of the authors. 

\bibliographystyle{IEEEtran}
\bibliography{Edgecomputing}

\end{document}